\documentclass{IEEEtran4PSCC}
\ifCLASSINFOpdf
   \usepackage[pdftex]{graphicx}
\else
   \usepackage[dvips]{graphicx}
\fi
\usepackage[cmex10]{amsmath}

\usepackage{enumitem}
\usepackage{multirow}
\usepackage{booktabs}
\usepackage{array}
\usepackage{cite}
\usepackage[hidelinks]{hyperref}
\usepackage{xurl}

\makeatletter
\let\old@ps@headings\ps@headings
\let\old@ps@IEEEtitlepagestyle\ps@IEEEtitlepagestyle
\def\psccfooter#1{%
    \def\ps@headings{%
        \old@ps@headings%
        \def\@oddfoot{\strut\hfill#1\hfill\strut}%
        \def\@evenfoot{\strut\hfill#1\hfill\strut}%
    }%
    \def\ps@IEEEtitlepagestyle{%
        \old@ps@IEEEtitlepagestyle%
        \def\@oddfoot{\strut\hfill#1\hfill\strut}%
        \def\@evenfoot{\strut\hfill#1\hfill\strut}%
    }%
    \ps@headings%
}
\makeatother

\psccfooter{%
        \parbox{\textwidth}{\hrulefill \\ \small{23rd Power Systems Computation Conference} \hfill \begin{minipage}{0.2\textwidth}\centering \vspace*{4pt} \includegraphics[scale=0.06]{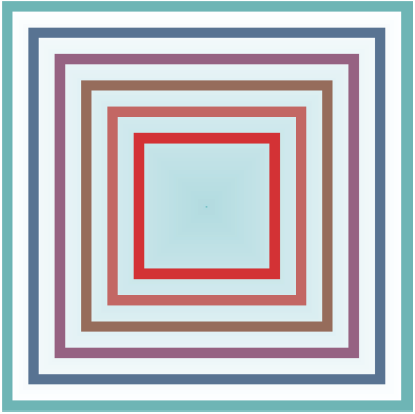}\\\small{PSCC 2024} \end{minipage} \hfill \small{Paris, France --- June 4 -- 7, 2024}}%
}

\begin{document}
%
\title{On the Financial Consequences of Simplified Battery Sizing Models without Considering Operational Details}

\author{\IEEEauthorblockN{Nam Trong Dinh\IEEEauthorrefmark{1},
Sahand Karimi-Arpanahi\IEEEauthorrefmark{1},
S. Ali Pourmousavi\IEEEauthorrefmark{1}, 
Mingyu Guo\IEEEauthorrefmark{2}, \\ Julian Lemos-Vinascco\IEEEauthorrefmark{3} and
Jon A. R. Liisberg\IEEEauthorrefmark{3}}
\IEEEauthorblockA{\IEEEauthorrefmark{1} School of Electrical and Mechanical Engineering\\
The University of Adelaide,
Adelaide, Australia}
\IEEEauthorblockA{\IEEEauthorrefmark{2} School of Computer and Mathematical Sciences\\
The University of Adelaide,
Adelaide, Australia}
\IEEEauthorblockA{\IEEEauthorrefmark{3} Watts A/S,
Køge, Denmark}
}

\maketitle

\begin{abstract}
Optimal battery sizing studies tend to overly simplify the practical aspects of battery operation within the battery sizing framework. Such assumptions may lead to a suboptimal battery capacity, resulting in significant financial losses for a battery project that could last more than a decade. In this paper, we compare the most common existing sizing methods in the literature with a battery sizing model that incorporates the practical operation of a battery, that is, receding horizon operation. Consequently, we quantify the financial losses caused by the suboptimal capacities obtained by these models for a realistic case study related to community battery storage (CBS). We develop the case study by constructing a mathematical framework for the CBS and local end users. Our results show that existing sizing methods can lead to financial losses of up to 22\%.
\end{abstract}


\thanksto{\noindent This project is supported by the Australian Government Research Training Program (RTP) through the University of Adelaide, and a supplementary scholarship provided by Watts A/S, Denmark.}

\section{Introduction}
\label{sec:introduction}

Community battery storage (CBS) has been recognised as a desirable solution for behind-the-meter (BTM) generation and demand management both in practice and in the literature \cite{Alkimos,Yarra,AusgridCBS,medi2020an,Ransan2021applying,dinh2022optimal}. In Australia, the trial of multiple CBS projects has led many distribution network service providers (DNSPs) to design new network tariffs specifically for CBS and end users within the neighbourhood \cite{AERtrial,Ausgridtrial}. These schemes generally incentivise the local use of the system (LUoS) for CBS located in low-voltage (LV) networks. Since the new CBS tariffs are designed to attract profit-making entities in the coming years \cite{subthreshold,Ausgridtrial}, it is crucial to accurately size the CBS to maximise the profit of the CBS owner.

In recent years, many research studies have been published on battery storage sizing \cite{khezri2022optimal}. However, the existing sizing models in the literature do not consider the practical aspects of battery operation. In these studies, a common approach is to assume a perfect prediction of power system parameters, e.g., electricity prices, renewable generation, and power consumption, to solve the planning problem over the entire sizing horizon, e.g., one year \cite{kani2020improving,tilman2021renewable,da2022a,khezri2021a}. These models guarantee a fast solution and can be scaled up for longer planning horizons, e.g., ten years. However, in practice, perfect knowledge of the future, even a couple of hours ahead, is impossible due to inherent uncertainties; hence, the battery operator's decisions can only be made under imperfect forecasts over a limited horizon, e.g., 24 hours ahead. 
To resolve this issue, Baker et al. \cite{baker2017energy} proposed an energy storage sizing formulation considering receding horizon operation (RHO) for battery units.
In this operational concept, the operator solves battery optimisation by predicting the parameters of the power system for a specified look-ahead horizon. The operator then only commits to the optimised solutions in the first interval of each receding horizon, while the remaining intervals are to ensure that the optimisation is not myopic. The optimisation problem must then be solved consecutively for the next receding horizons as new forecast data becomes available, a process that can be time- and resource-intensive. To address this issue, the authors in \cite{baker2017energy, taheri2019energy, moghaddam2019optimal} coupled all receding horizons together, rather than solving the optimisation problem sequentially, to solve them simultaneously as a single optimisation problem. However, this coupling approach can negatively affect the optimal solutions because one horizon can be strongly influenced by many subsequent shifted horizons. Also, this is not the way a battery unit operates in practice. In general, these methods can lead to suboptimal battery capacity, the impact of which has not been highlighted or quantified compared to the global optimal solution. In this context, this paper offers two main contributions:
\vspace{-0.3em}
\begin{itemize}[wide]
    \item Quantifying the financial losses from suboptimal battery capacities obtained by the existing battery sizing methods and providing structural insights on the behaviour of these models.
    \item Implementing a mathematical RHO framework for local end users and CBS operators by considering realistic network charges, trialled by the DNSPs that promote LUoS. 
\end{itemize}

\section{Local market model with CBS}
Current CBS trials in Australia allow solar end users (prosumers) within a local neighbourhood to virtually store their excess solar photovoltaic (PV) generation to CBS in exchange for solar credits \cite{Alkimos,Yarra,AusgridCBS}. At night, prosumers can use solar credits to offset their consumption. Additionally, the CBS operator collaborates with an existing electricity retailer to propose a time-of-use tariff structure with a high tariff during peak demand hours to promote a price-based demand response. The high peak demand tariff directs local prosumers to use their solar credits to offset consumption during those hours, typically in the early evening. Thus, the CBS operator automatically offsets prosumers' usage in the first instance of peak-demand hours. Nonetheless, there are a few retailers, e.g., Amber Electric \cite{AmberElectric}, that allow residential customers to pay for electricity at real-time (RT) wholesale spot prices. In this way, prosumers should be free to decide when to offset their consumption depending on the RT prices and their available solar credits. The fluctuation of RT prices also encourages prosumers to practice demand response to minimise their electricity bill \cite{smart2022the}. Therefore, in this paper, we adopt the price-responsive model from \cite{dinh2022optimal, dinh2023modelling} to model end-user behaviour under fluctuations in RT prices, while also considering realistic network usage charges from DNSPs when storing electricity to CBS and consuming energy from CBS. 
To accurately model the problem under an imperfect scenario, we solve the optimisation models
using 30-minute resolution pre-dispatch (PD) prices of the Australian National Electricity Market (NEM) \cite{aemowebsite}. These PD spot prices, determined by aggregate demand forecasts and generator bids, are updated every half hour. Using NEM PD prices, we update the forecast of electricity prices at each receding horizon. 

\subsection{End-user Model}
\label{sec:pro_model}
\begin{figure}[!h]
    \centerline{\includegraphics[trim={2.1cm 11.0cm 3.1cm 11.2cm}, clip, width=\linewidth]{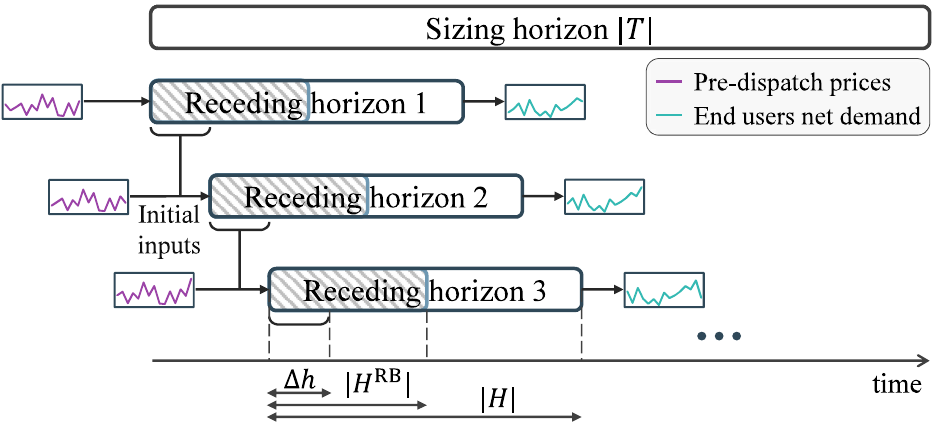}}
    \vspace{-0.7em}
    \caption{A flowchart showing end users RHO framework}
    \label{fig:end_user_rho}
    \vspace{-0.3em}
\end{figure}

While PD prices offer continuously updated electricity price forecasts, there has not yet been a dataset for continuously updated residential price-responsive consumption forecasts. To this end, we solve the end-user optimisation problem in an RHO framework to obtain the dynamic consumption behaviour for each receding horizon. The end-user RHO model can be seen in Fig. \ref{fig:end_user_rho}.

We denote the set of end users by $N = \{1,2,\dots,|N|\}$. In RHO, the optimisation is solved each time on a receding horizon that includes the set of time intervals $H = \{1,2,\cdots,|H|\}$, where $|H|$ is the number of intervals in each horizon, and the optimisation recedes one interval each time. Overall, the optimisation is solved $|T|$ times for all receding horizons, denoted by $j \in J = \{1,2,\cdots,|T|\}$, where $|T|$ represents the length of the battery sizing horizon. For each receding horizon $j$, the end-user optimisation can be formulated as follows:
\begin{subequations}
\label{eqn:pro_sol}
\begin{flalign}
\label{eqn:max_uti_pro}
    \min_{\mathbf{\Psi^{\text{user}}_{n,j}}} O^{\text{user}}_{n,j} = \sum_{h \in H} & \bigg[\lambda^{\text{PD}}_{j,h} x^{\text{g}}_{n,j,h} + \lambda^{\text{ExP}}_{j,h} x^-_{n,j,h} + \lambda^{\text{ImP}}_{j,h} x^+_{n,j,h} \nonumber \\
    & + \! \frac{1 + \tau_n \! \cdot \! h \! \cdot \! \kappa_n}{1 + h \! \cdot \! \kappa_n} D(x_{n,j,h}) \bigg] \quad \forall n \! \in \! N, &&
\end{flalign}
where
\begin{flalign}
\label{eqn:pro_discomfort}
    D(x_{n,j,h})\! = \! -\lambda^{\text{PD}}_{\text{max},j}  \left(1 \!+\! \frac{x_{n,j,h} \!-\! \hat{x}_{n,j,h}}{2\beta_{n,j,h}\hat{x}_{n,j,h}}\right) (x_{n,j,h}\! -\! \hat{x}_{n,j,h}) &&
\end{flalign}
s.t.
\begin{flalign}
\label{eqn:rebound_const}
    \sum_{h \in H^\text{RB}} \! x_{n,j,h} = \! \sum_{h \in H^\text{RB}} \! \hat{x}_{n,j,h} + \Delta x_{n,j} \quad \forall n \in N, &&
\end{flalign}
\begin{flalign}
\label{eqn:cons_boundary}
    \underline{x}_{n,j,h} \leq x_{n,j,h} \leq \overline{x}_{n,j,h} \quad \forall n \in N, \; \forall h \in H, &&
\end{flalign}
\begin{flalign}
\label{eqn:prosumer_net}
    x_{n,j,h} - G^\text{u}_{n,j,h} = x^+_{n,j,h} - x^-_{n,j,h} \quad \forall n \in N, \; \forall h \in H, &&
\end{flalign}
\begin{flalign}
\label{eqn:complement_net_restriction}
    0 \leq x^+_{n,j,h} \perp x^-_{n,j,h} \geq 0 \quad \forall n \in N, \; \forall h \in H, &&
\end{flalign}
\begin{flalign}
\label{eqn:used_solar}
    G^\text{u}_{n,j,h} \leq G_{n,j,h} \quad \forall n \in N, \; \forall h \in H, &&
\end{flalign}
\begin{flalign}
\label{eqn:credit_offset}
    x^+_{n,j,h} - \delta_{n,j,h} = x^{\text{g}}_{n,j,h} \quad \forall n \in N, \; \forall h \in H, &&
\end{flalign}
\begin{flalign}
\label{eqn:cumulative_credit}
    C_{n,j,h} = C^{\text{init}}_{n,j} + \! \sum_{l=1}^h (x^-_{n,j,l} \! - \! \delta_{n,j,l}) \quad \forall n \in \! N, \; \forall h \in \! H, &&
\end{flalign}
\begin{flalign}
\label{eqn:pro_sign_cons}
    x_{n,j,h}, x^+_{n,j,h}, x^-_{n,j,h}, G^\text{u}_{n,j,h}, \delta_{n,j,h}, & \, x^{\text{g}}_{n,j,h}, C_{n,j,h} \geq 0 \nonumber \\ 
    & \;\; \forall n \in N, \; \forall h \in H, &&
\end{flalign}
\end{subequations}
where $\mathbf{\Psi^{\text{user}}_{n,j}} = \{x_{n,j,h}, x^+_{n,j,h}, x^-_{n,j,h}, G^{\text{u}}_{n,j,h}, \delta_{n,j,h}, x^{\text{g}}_{n,j,h},$ $C_{n,j,h}\}$. 
As seen in \eqref{eqn:max_uti_pro}, end users want to minimise their electricity cost and the discomfort caused by load shifting. The proumers' electricity cost consists of the energy payment at PD prices, $\lambda^{\text{PD}}_{j,h}$, for consumption from the grid, $x^{\text{g}}_{n,j,h}$, 
and the network usage charges, i.e., $\lambda^{\text{ExP}}_{j,h}$ and $\lambda^{\text{ImP}}_{j,h}$, for exported, $x^-_{n,j,h}$, and imported (consumed) electricity, $x^+_{n,j,h}$, respectively. Here, the network usage charges are set by the DNSPs. The (dis)comfort model is integrated with the time inconsistency and loss aversion properties of behavioural economics as introduced in \cite{dinh2023modelling}. In particular, the time inconsistency is represented by the fraction in the last term of \eqref{eqn:max_uti_pro}, which depends on the degree of short-term discounting, $\kappa_n$, and the degree of long-term discounting, $\tau_n$. On the other hand, the loss aversion is modelled by a quadratic function in \eqref{eqn:pro_discomfort} that depends on actual consumption, $x_{n,j,h}$, expected consumption, $\hat{x}_{n,j,h}$, price elasticity, $\beta_{n,j,h}$, and price reference, $\lambda^{\text{PD}}_{\text{max},j} := \max\{\lambda^{\text{PD}}_{j,h}|h \in H\}$, which is adopted from \cite{dinh2023modelling}. 
Constraint \eqref{eqn:rebound_const} ensures that demand response is only provided by load shifting such that, in each receding horizon, the sum of actual consumption remains the same as the total expected consumption and the consumption deviation from the previous receding horizons, $\Delta x_{n,j}$. To make the model realistic, unlike existing studies, e.g., \cite{werner2021pricing,dinh2022optimal}, we enforce the rebound effect of shiftable loads to occur within the first few hours rather than the whole receding horizon. Therefore, $H^{\text{RB}}$ denotes the rebound horizon such that $H^{\text{RB}} \subset H$. Constraint \eqref{eqn:cons_boundary} sets the lower, $\underline{x}_{n,j,h}$, and upper, $\overline{x}_{n,j,h}$, bounds of consumption in each interval. 
Constraint \eqref{eqn:prosumer_net} separates the net demand into exported and imported (consumed) electricity and restricts them from simultaneously having non-zero values through the complementarity constraint in \eqref{eqn:complement_net_restriction}. In Australia, renewable energy constitutes a large portion of the energy mix. This has often led to wholesale prices dropping below zero, sometimes reaching as low as $-$\$1000/MWh \cite{aemofactsheet}. As a result, the optimal strategy during these intervals is to curtail solar generation. To do this, we consider `used' solar energy, $G^\text{u}_{n,j,h}$, in \eqref{eqn:prosumer_net} and constrain it in \eqref{eqn:used_solar}. Constraint \eqref{eqn:credit_offset} determines the consumption from the utility grid, $x^{\text{g}}_{n,j,h}$, after deducting the solar credits used, $\delta_{n,j,h}$. Constraint \eqref{eqn:cumulative_credit} determines the cumulative solar credits over time, $C_{n,j,h}$, with $C^{\text{init}}_{n,j}$ denoting the initial cumulative solar credits in each receding horizon. Due to the sequential solving of the RHO, we have $C^{\text{init}}_{n,j} \!=\! C^\star_{n,j-1}$ and $\Delta x_{n,j} \!=\! \sum_{l=1}^j (\hat{x}^\star_{n,l} - x^\star_{n,l})$ as parameters determined by the previous receding horizons. Here, we define the variables with ($\star$) as the optimised values committed from previous receding horizons. Lastly, we define the sign of the variables in \eqref{eqn:pro_sign_cons}.

\subsection{CBS Operation}
\label{sec:CBS_oper}

\begin{figure}[!h]
    \centerline{\includegraphics[trim={2.1cm 11.2cm 3.1cm 11.2cm}, clip, width=\linewidth]{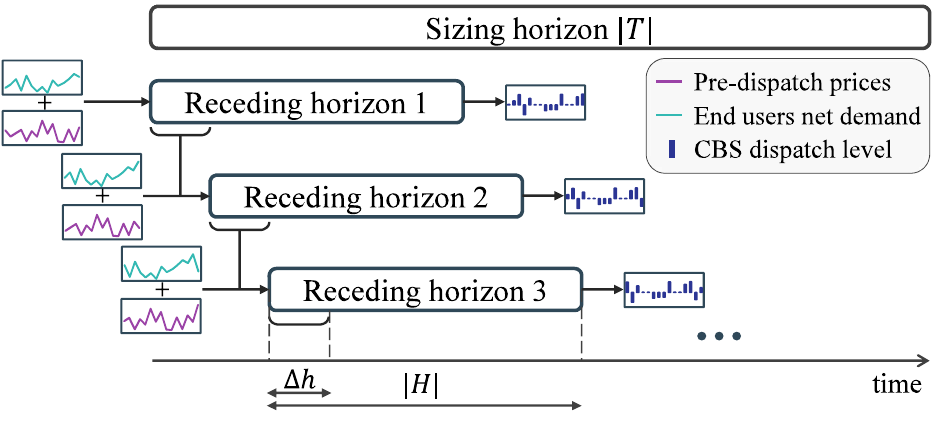}}
    \vspace{-0.7em}
    \caption{A flowchart showing the CBS RHO framework}
    \label{fig:cbs_rho}
    \vspace{-0.3em}
\end{figure}

The solution to the end-user optimisation problem is the changing consumption behaviour over time. As a result, in the CBS optimisation problem, the uncertain parameter is not only the prices but also the consumption of end users, as shown in Fig. \ref{fig:cbs_rho}. For each receding horizon $j$, the optimisation problem for the CBS is as follows:
\begin{subequations}
\label{eqn:CBS_operation}
\begin{flalign}
\label{eqn:CBS_profit}
    \min_{\mathbf{\Psi^{\text{CBS}}_j}} O^{\text{CBS}}_j = & \sum_{h \in H} \big( \lambda^{\text{PD}}_{j,h} \vartheta^+_{j,h} + \lambda^{\text{g}} \vartheta^{\text{g}}_{j,h} + \lambda^\text{ThP} P^\text{dc}_{j,h} \Delta h \big) \nonumber \\
    & - \lambda^\text{peak}(\Upsilon_j^\text{local} - \Upsilon_j^\text{user}) &&
\end{flalign}
s.t.
\begin{flalign}
\label{eqn:net_demand}
    \sum_{n \in N} \!\! \big( x^+_{n,j,h} \! - \! x^-_{n,j,h} \big) \! + \! P_{j,h} \Delta h = \vartheta^+_{j,h} \! - \! \vartheta^-_{j,h} \quad \forall h \! \in \! H, &&
\end{flalign}
\begin{flalign}
\label{eqn:comeplement_net_demand}
    0 \leq \vartheta^+_{j,h} \perp \vartheta^-_{j,h} \geq 0 \quad \forall h \in H, &&
\end{flalign}
\begin{flalign}
\label{eqn:peak_demand}
    \Upsilon^\text{local}_j \geq \max \left( \vartheta^+_{j,h}, \max_{l \in \{1,2,\dots,j\}} \vartheta^{+\star}_l \right) \quad \forall h \in H, &&
\end{flalign}
\begin{flalign}
\label{eqn:battery_soc}
    E_{j,h} = E^{\text{init}}_j + \sum_{l=1}^h \big( P^{\text{ch}}_{j,l} - \frac{1}{\Gamma} P^{\text{dc}}_{j,l}\big) \Delta h \quad \forall h \in H, &&
\end{flalign}
\begin{flalign}
\label{eqn:charging_power}
    P_{j,h} = P^{\text{ch}}_{j,h} - P^{\text{dc}}_{j,h} \quad \forall h \in H, &&
\end{flalign}
\begin{flalign}
\label{eqn:charging_limit}
    -\frac{E^{\text{cap}}}{T^\text{c}} \leq P_{j,h} \leq \frac{E^{\text{cap}}}{T^\text{c}} \quad \forall h \in H, &&
\end{flalign}
\begin{flalign}
\label{eqn:soc_boundary}
    \underline{\text{SoC}} \, E^{\text{cap}} \leq E_{j,h} \leq \overline{\text{SoC}} \, E^{\text{cap}} \quad \forall h \in H, &&
\end{flalign}
\begin{flalign}
\label{eqn:ending_SOC}
    E^{\text{init}}_j = E_{j,h=|H|}, &&
\end{flalign}
\begin{flalign}
\label{eqn:LUoS}
    P^{\text{ch}}_{j,h} \Delta h - \sum_{n \in N} x^-_{n,j,h} \leq \vartheta^{\text{g}}_{j,h} \quad \forall h \in H, &&
\end{flalign}
\begin{flalign}
\label{eqn:ope_sign_cons_1}
    E_{j,h}, P^{\text{ch}}_{j,h}, P^{\text{dc}}_{j,h}, \vartheta^+_{j,h}, \vartheta^-_{j,h}, \vartheta^{\text{g}}_{j,h}, \Upsilon^\text{local}_j \geq 0 \quad \forall h \in H, &&
\end{flalign}
\end{subequations}
where $\mathbf{\Psi^{\text{CBS}}_j} \!=\!\! \{ E_{j,h},\! P^{\text{ch}}_{j,h},\! P^{\text{dc}}_{j,h},\! \vartheta^+_{j,h},\! \vartheta^-_{j,h},\! \vartheta^{\text{g}}_{j,h}, \! \Upsilon^\text{local}_j\!\}$. Similarly to the end-user problem, the CBS operation is solved sequentially with the initial state-of-charge (SoC), $E^{\text{init}}_j = E^\star_{j-1}$, passed from the previous horizon as initial state input, in which the initial value (i.e., $j \! = \! 1$) is set to zero. The objective of the CBS operator in \eqref{eqn:CBS_profit} is to minimise the net cost. This cost includes the energy payment to the wholesale market, the network usage charge, $\lambda^{\text{g}}$, when CBS is charged through imported electricity from the utility grid, $\vartheta^\text{g}_{j,h}$, and the CBS throughput charge, $\lambda^\text{ThP}$, on CBS discharging power to prevent multiple CBS charge and discharge cycles in each receding horizon. The last term in \eqref{eqn:CBS_profit} depicts the revenue from peak demand reduction which is the difference between the end users' peak demand, $\Upsilon^\text{user}_j = \max \{ \sum_{n \in N} (x^+_{n,j,h} - x^-_{n,j,h}| h \in H) \}$, and the peak demand after considering the CBS operation, $\Upsilon^\text{local}_j$. Here, the network usage charge, $\lambda^\text{g}$, and peak demand charge, $\lambda^\text{peak}$, are set by the DNSPs. Constraint \eqref{eqn:net_demand} separates the net demand of the whole neighbourhood into imported, $\vartheta^+_{j,h}$, and exported, $\vartheta^-_{j,h}$, electricity. To restrict them from simultaneously taking non-zero values, complementarity constraints are implemented in \eqref{eqn:comeplement_net_demand}. We define the peak demand of the local neighbourhood in \eqref{eqn:peak_demand}, which considers both the potential maximum net demand in the look-ahead horizon and the observed peak demand in previous receding horizons. Constraint \eqref{eqn:battery_soc} represents the evolution of the CBS SoC over time, where the charging, $P^\text{ch}_{j,h}$, and discharging, $P^\text{dc}_{j,h}$, power are depicted in \eqref{eqn:charging_power}. Moreover, in \eqref{eqn:battery_soc}, $\Gamma$ represents the CBS round-trip efficiency, and $\Delta h$ represents the granularity of the intervals. Constraint \eqref{eqn:charging_limit} limits the CBS (dis)charging power with respect to the CBS capacity and battery duration, $T^\text{c}$. Constraint \eqref{eqn:soc_boundary} limits the CBS SoC within the lower, $\underline{\text{SoC}}$, and upper, $\overline{\text{SoC}}$, bounds. To avoid fully discharging at the end of each receding horizon, constraint \eqref{eqn:ending_SOC} sets the ending SoC equal to the initial SoC. Constraint \eqref{eqn:LUoS} determines the imported electricity from the utility grid for CBS charging activity. As mentioned in section \ref{sec:introduction}, the trial CBS tariffs promote the LUoS. Thus, there is no cost when charging the CBS using the excess PV generation within the local neighbourhood. In contrast, the CBS operator must pay a fixed charge, $\lambda^\text{g}$, when charging from the utility grid. Lastly, \eqref{eqn:ope_sign_cons_1} represents the sign of the variables. Note that in the CBS operation, the CBS capacity, $E^\text{cap}$, is a known parameter.

\subsection{Ground Truth Cost Calculation}

Since we operate the CBS in \eqref{eqn:CBS_operation} with respect to the PD prices, it is necessary to calculate the ground truth cost of the CBS operation by applying the optimised solutions committed from \eqref{eqn:CBS_operation} to the RT dispatch (cleared) prices, $\lambda^\text{RT}$. Additionally, in \eqref{eqn:CBS_profit}, we assess the revenue from peak demand reduction for each receding horizon separately. However, in practice, DNSPs typically assess peak demand on a yearly basis. As a result, we calculate the ground truth cost as follows:
\begin{flalign}
\label{eqn:ground_truth}
    \text{Total cost} = & \sum_{j \in J} \left( \lambda^{\text{RT}}_{j} \vartheta^{+\star}_{j} + \lambda^{\text{g}} \vartheta^{\text{g}\star}_{j} + \lambda^\text{ThP} P^{\text{dc}\star}_j \Delta h \right) \nonumber \\
    & - \lambda^\text{peak} (\Upsilon^{\text{local}\star} - \Upsilon^{\text{user}\star}) + \lambda^\text{bat} E^\text{cap}, &&
\end{flalign}
where $\Upsilon^{\text{local}\star} = \max \{ \vartheta^{+\star}_{j} | j \in J \}$ and $\Upsilon^{\text{user}\star} = \max \{\sum_{n \in N}$ $(x^{+\star}_{n,j} - x^{-\star}_{n,j}) | j  \in \! J \}$. We also include the cost of CBS, $\lambda^\text{bat} E^\text{cap}$, as part of the total project cost. Here, we only consider the net cost related to the CBS and its operation to give an accurate comparison among different sizing methods, as introduced in the subsequent section. 

\section{Battery sizing methods}
\label{sec:battery_sizing}

\subsection{Exhaustive Search for (\textit{Exact}) Battery Sizing}
Since the CBS operates under the RHO regime, we cannot obtain the optimal CBS capacity in one single optimisation. Instead, the lowest project cost in \eqref{eqn:ground_truth} must be determined by examining different values of battery capacity, $E^\text{cap}$, for CBS operation in \eqref{eqn:CBS_operation}. Therefore, in this paper, we iteratively assess all possible CBS capacities with a step of 5 kWh to find the global optimal value.

\subsection{Without Receding Horizon (\textit{W/o RH})}
\label{sec:w/oRH}
As mentioned in section \ref{sec:introduction}, a common battery sizing approach is to assume a perfect prediction of uncertain parameters, i.e., electricity prices and prosumers' consumption in our case, 
and solve a planning problem over the entire sizing horizon. 
To size the CBS without RHO, we solved a modified version of \eqref{eqn:CBS_operation}, where instead of looking at all the intervals in $H$, we only consider the first interval in $H$, i.e., $h=1$, and sum over $j \in J$ in \eqref{eqn:CBS_profit}. Moreover, we remove \eqref{eqn:ending_SOC} since now there is only one ending SoC. We formulate the battery sizing model for the \textit{W/o RH} method as follows:
\begin{subequations}
\label{eqn:CBS_sizing_w/oRH}
\begin{flalign}
\label{eqn:CBS_profit_sizing}
    \min_{\mathbf{\Psi^{\text{WoRH}}}} S^\text{WoRH} = & \sum_{j \in J} \big(\lambda^{\text{RT}}_{j} \vartheta^+_{j} + \lambda^{\text{g}} \vartheta^{\text{g}}_{j}  + \lambda^\text{ThP} P^{\text{dc}}_{j} \Delta h )\nonumber \\
    & - \lambda^\text{peak} (\Upsilon^\text{local} - \Upsilon^{\text{user}\star}) + \lambda^\text{bat} E^\text{cap}, &&
\end{flalign}
\begin{flalign}
\label{eqn:s.t.sizing_without_RHO}
\text{s.t. } \eqref{eqn:net_demand}\text{--}\eqref{eqn:comeplement_net_demand}, \eqref{eqn:battery_soc}\text{--}\eqref{eqn:soc_boundary}, \eqref{eqn:LUoS} \text{ such that } h=1, \; \forall j \in J, &&
\end{flalign}
\begin{flalign}
\label{eqn:peak_demand_woRH}
    \Upsilon^\text{local} \geq \vartheta^+_{j} \quad \forall j \in J, &&
\end{flalign}
\begin{flalign}
\label{eqn:sign_cons_woRH}
    E_{j}, P^{\text{ch}}_{j}, P^{\text{dc}}_{j}, \vartheta^+_{j}, \vartheta^-_{j}, \vartheta^{\text{g}}_{j}, \Upsilon^\text{local}, E^\text{cap} \geq 0 \quad \forall j \in J, &&
\end{flalign}
\end{subequations}
where $\mathbf{\Psi^{\text{WoRH}}} = \{ E_{j}, P^{\text{ch}}_{j}, P^{\text{dc}}_{j}, \vartheta^+_{j}, \vartheta^-_{j}, \vartheta^{\text{g}}_{j}, \Upsilon^\text{local}, E^\text{cap}\}$. As $E^\text{cap}$ is a decision variable in this optimisation, we can directly obtain the CBS capacity in a single optimisation process. Note that we have normalised both $\lambda^\text{peak}$ and $\lambda^\text{cap}$ to align with the length of the receding horizons. In \eqref{eqn:CBS_profit_sizing}, we use the RT prices, representing a scenario with perfect foresight. To explore the impact of forecast prices, we can replace them with PD prices. Particularly, in this paper, we conduct two scenarios for the W/o RH sizing method: one with RT prices and another with 30-minute look-ahead PD prices, i.e., PD prices at $h=1$.

\subsection{Coupled Receding Horizons (\textit{Coupled RH})}
\label{sec:coupled_RH}
In \cite{baker2017energy}, the battery sizing and operation are optimised simultaneously considering all receding horizons in one optimisation problem. Therefore, to formulate this problem based on the model in \eqref{eqn:CBS_operation}, we introduce a new constraint, $E^{\text{init}}_j = E_{j-1,h=1}$, which was originally used to set the initial state parameter in each receding horizon of the CBS operation. We formulate the battery sizing model for the \textit{Coupled RH} approach as follows:
\begin{subequations}
\label{eqn:CBS_sizing_coupling}
\begin{flalign}
\label{eqn:CBS_sizing_coupled}
    \min_{\mathbf{\Psi^{\text{CoRH}}}} S^{\text{CoRH}} & \! = \! \sum_{\omega \in \Omega} \bigg[ \frac{1}{|H|} \sum_{j \in J_\omega} \sum_{h \in H} \big( \lambda^{\text{PD}}_{j,h} \vartheta^+_{j,h} + \lambda^\text{ThP} P^\text{dc}_{j,h} \Delta h \nonumber \\
    & \!\! + \! \lambda^{\text{g}} \vartheta^{\text{g}}_{j,h} \big) \! \bigg] \!\! - \! \lambda^\text{peak}(\Upsilon^\text{local} \! - \! \Upsilon^{\text{user}\star}) \! + \! \lambda^\text{bat} E^\text{cap} &&
\end{flalign}
\begin{flalign}
\label{eqn:s.t.sizing_coupling}
\text{s.t. } \eqref{eqn:net_demand}\text{--}\eqref{eqn:ope_sign_cons_1} \quad \forall j \in J_\omega, \; \forall \omega \in \Omega, &&
\end{flalign}
\begin{flalign}
\label{eqn:initial_SOC}
E^{\text{init}}_j = E_{j-1,h=1} \quad \forall j \in J_\omega \setminus \{1\}, \; \forall \omega \in \Omega, &&
\end{flalign}
\begin{flalign}
\label{eqn:sign_cons_CoRH}
    E_{j,h}, P^{\text{ch}}_{j,h}, P^{\text{dc}}_{j,h}, \vartheta^+_{j,h}, \vartheta^-_{j,h}, & \; \vartheta^{\text{g}}_{j,h}, \Upsilon^\text{local}, E^\text{cap} \geq 0 \nonumber \\
    & \forall h \in H, \; \forall j \in J_\omega, \; \forall \omega \in \Omega &&
\end{flalign}
\end{subequations}
where $\mathbf{\Psi^{\text{CoRH}}} = \{ E_{j,h}, P^{\text{ch}}_{j,h}, P^{\text{dc}}_{j,h}, \vartheta^+_{j,h}, \vartheta^-_{j,h}, \vartheta^{\text{g}}_{j,h}, \Upsilon^\text{local}, E^\text{cap}\}$. Similar to the W/o RH method, the optimal $E^\text{cap}$ in Coupled RH approach can be obtained in one single optimisation process. Since all receding horizons are considered simultaneously, we need to divide the extended battery sizing duration into smaller periods $\omega \in \Omega$, where each period contains $|J_\omega|$ receding horizons. This is done to avoid one receding horizon from looking too far ahead into the future. For example, in \cite{baker2017energy}, each receding horizon was 1-hour long with 10-minute granularity, equivalent to $|H| = 6$ in our model. Additionally, their sizing model considered each period to be one full day, i.e., $J_\omega = 144$. In contrast, our model considers close to one-day look-ahead with 30-minute intervals. As a result, we set each period $\omega$ to one week in our study. This time frame allows the sizing model enough flexibility to establish the RHO without allowing receding horizons looking far into the future to interfere with current calculations. Lastly, due to the coupled receding horizons, we need to take the weighted sum of all the receding horizons by dividing over $|H|$ as seen in \eqref{eqn:CBS_sizing_coupled}.

\section{Numerical Study}

\subsection{Simulation Setup}

\begin{table}[t]
    \caption{Network usage charges and end user elasticity at different time intervals}
    \vspace{-0.2 em}
    \centering
    \label{tab:prosumer_charges}
    \begin{tabular}{c|ccc}
    \specialrule{.15em}{0em}{0.08em}
    \textbf{Interval} & $\lambda^\text{ExP}_{j,h}$ (\textcent/kWh) & $\lambda^\text{ImP}_{j,h}$ (\textcent/kWh) & $\beta_{n,j,h}$ \\
    \specialrule{.08em}{0.08em}{0em}
    \textbf{01:00--05:00} & 0 & 3.3095 & $[-0.2, \, -0.3]$ \\
    \textbf{05:00--10:00} & 0 & 3.3095 & $[-0.3, \, -0.5]$ \\
    \textbf{10:00--14:00} & 1.8500 & 3.3095 & $[-0.3, \, -0.5]$ \\
    \textbf{14:00--20:00} & $-$27.7957 & 27.7957 & $[-0.5, \, -0.7]$ \\
    \textbf{20:00--01:00} & 0 & 3.3095 & $[-0.3, \, -0.5]$ \\
    \specialrule{.1em}{0em}{0em}
    \end{tabular}
    \vspace{-1.0 em}
\end{table}

\begin{table}[t]
  \caption{CBS data and simulation input parameters}
  \vspace{-0.2 em}
  \label{tab:simu_input}
  \centering
  \begin{tabular}{p{0.1\textwidth}r|p{0.1\textwidth}r}
  \specialrule{.15em}{0em}{0.2em} 
  \multicolumn{2}{l}{\textbf{Battery data}} & \multicolumn{2}{l}{\textbf{Other simulation parameters}} \\
  \specialrule{.08em}{0em}{0em}
  $\Gamma$ & $90$\% & $\underline{x}_{n,j,h}$ & $0.5\hat{x}_{n,j,h}$ \\
  $\underline{\text{SoC}}$, $\overline{\text{SoC}}$ & $0$\%, $100$\% & $\overline{x}_{n,j,h}$ & $1.5\hat{x}_{n,j,h}$ \\
  $\Delta h$ & 0.5 hours & $|H^{\text{RB}}|$ & 12 (6 hours) \\
  $T^\text{c}$ & $2$ hours & $|H|$ & 32 (16 hours) \\
  $\lambda^\text{bat}$ & \$800/kWh & $\kappa_n$ & $[0.1, \, 0.5]$ \\
  $\lambda^\text{ThP}$ & 3.2 \textcent/kWh & $\tau_n$ & 0.2 \\
  $\lambda^\text{g}$ & 1.61 \textcent/kWh & $\lambda^\text{peak}$ & 0.33 \$/kW $\!\! \cdot \!\!$ year \\
  \specialrule{.1em}{0em}{0em}
  \end{tabular}
  \vspace{-1.0 em}
\end{table}

\begin{itemize}[wide]
    \item \textbf{End user profiles} were collected from the Solar Home dataset with 60 solar prosumers and 60 non-solar consumers in New South Wales (NSW), Australia \cite{Ausgrid2012data}. The dataset contains half-hourly electricity consumption and gross rooftop PV generation in 2012. Due to the increase in rooftop PV capacity in recent years \cite{AEC2021}, we uniformly scaled up the PV generation profiles of all prosumers by three times, giving an average rooftop PV capacity of 5.1 kWp.
    \item \textbf{Electricity prices and network charges} were collected for the NSW region in 2021 \cite{aemowebsite}. The new network charges for end users, recently tested in that region, and the CBS tariff were collected from the DNSP in NSW \cite{Ausgridtrial}. Table \ref{tab:prosumer_charges} shows the end-user network usage charges at different times of the day, where the negative value of $\lambda^\text{ExP}$ represents an export reward for rooftop solar PV energy.
    \item \textbf{CBS data and simulation parameters} are summarised in Table \ref{tab:simu_input}. At any time of the day, PD prices in the NEM are available for a minimum of 16 hours ahead \cite{aemowebsite}. Thus, in our optimisation, the length of one receding horizon is considered to be 16 hours. The price elasticity of electricity demand is relatively low and varies depending on the time of the day. To capture this, we randomly generated a price elasticity value, $\beta_{n,j,h}$, for each end user from a uniform distribution based on the time of day. Table \ref{tab:prosumer_charges} shows the time bands and the ranges of the distribution.
    \item \textbf{Simulation period} was divided into in-sample and out-of-sample periods. The in-sample period, comprising the second week of each month (84 days in total) of the selected years, is used to size the CBS capacity. In contrast, the out-of-sample period, which includes the third week of each month (also 84 days), is used to evaluate and affirm the performance of the different sizing methods in section \ref{sec:battery_sizing}. The code and data used for our simulations can be accessed at \cite{githubproject}.
\end{itemize}

\subsection{Simulation Results and Discussions}
\subsubsection{Pre-dispatch prices and end user consumption}

\begin{figure}[!t]
    \centerline{\includegraphics[trim={0.25cm 0cm 0.2cm 0.2cm}, clip, width=\linewidth]{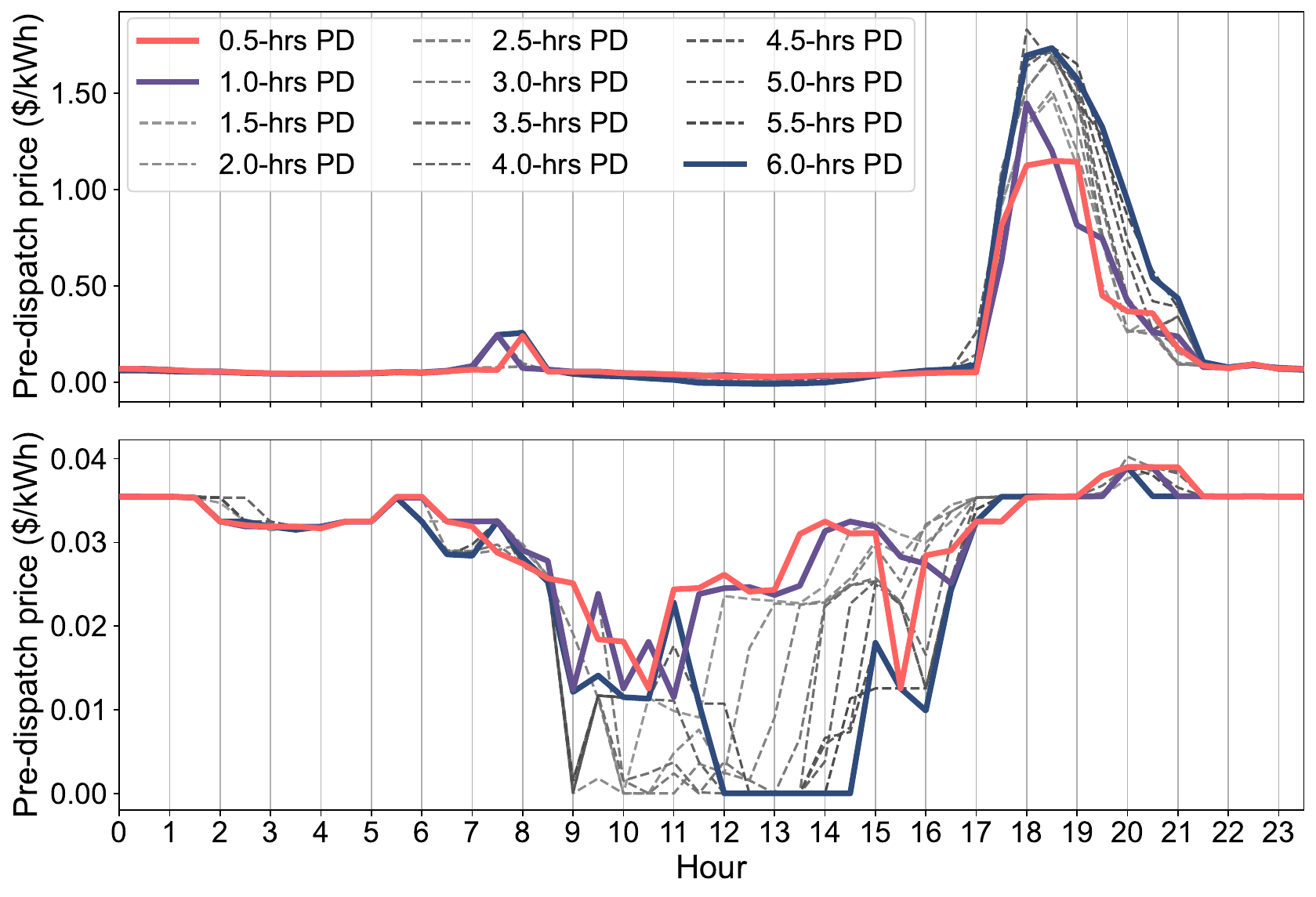}}
    \vspace{-0.3em}
    \caption{PD prices at different receding horizons. The upper figure depicts average PD prices from the in-sample period, while the lower figure shows the PD prices of a specific day in January}
    \label{fig:look-ahead_price}
    \vspace{-0.3em}
\end{figure}

\begin{figure}[t]
    \centerline{\includegraphics[trim={0.25cm 0cm 0.2cm 0.2cm}, clip, width=\linewidth]{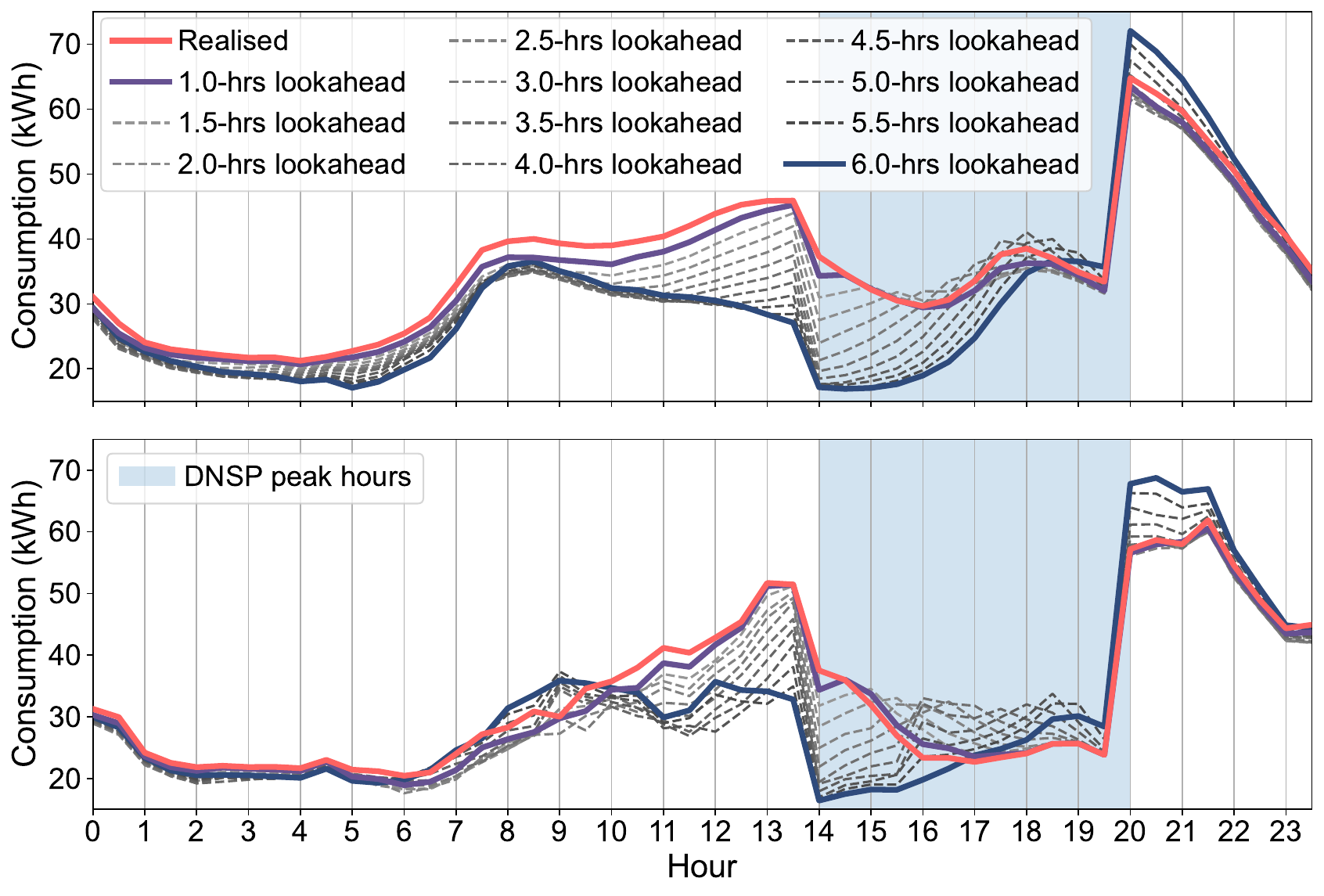}}
    \vspace{-0.3em}
    \caption{Realised and expected consumption at different receding horizons. The upper figure shows the average consumption from the in-sample period, while the lower figure shows the consumption of the same day as in Fig. \ref{fig:look-ahead_price}}
    \label{fig:look-ahead_cons}
    \vspace{-1.0em}
\end{figure}

\begin{table*}[!t]
  \caption{Optimal CBS capacity and normalised annual cost with daily average cycle for in-sample and out-of-sample periods}
  \vspace{0 em}
  \label{tab:optimal_va}
  \centering
  \setlength{\tabcolsep}{0.5em}
  \renewcommand{\arraystretch}{1.2}
  \begin{tabular}{c|c|cccc|cccc}
  \specialrule{.15em}{0em}{0em}
  & & \multicolumn{4}{c|}{\textbf{In-sample}} & \multicolumn{4}{c}{\textbf{Out-of-sample}} \\
  \cline{3-10}
  \multirow{-2}{*}{\textbf{Methods}} & \multirow{-2}{*}{\begin{tabular}{@{}c@{}}\textbf{Capacity}\\ \textbf{(kWh)}\end{tabular}} & \textbf{Energy charge} & \textbf{Peak reduction} & \textbf{Total cost} & \textbf{Avg cycle} & \textbf{Energy charge} & \textbf{Peak reduction} & \textbf{Total cost} & \textbf{Avg cycle} \\
  \specialrule{.08em}{0em}{0em}
  \textbf{Exact} & 250 & \$37.8k & \begin{tabular}{@{}c@{}}$-$\$13.8k \\ (114.1 kW)\end{tabular} & \begin{tabular}{@{}c@{}}\textbf{\$47.5k} \\ \textbf{(+$\,$0.0\%)}\end{tabular} & 1.01 & \$53.2k & \begin{tabular}{@{}c@{}}$-$\$6.2k \\ (50.8 kW)\end{tabular} & \begin{tabular}{@{}c@{}}\textbf{\$70.5k} \\ \textbf{(+$\,$0.0\%)}\end{tabular} & 1.04 \\
  \hline
  \begin{tabular}{@{}c@{}}\textbf{W/o RH}\\ \textbf{perfect foresight}\end{tabular} & 320 & \$36.3k & \begin{tabular}{@{}c@{}}$-$\$13.8k \\ (114.1 kW)\end{tabular} & \begin{tabular}{@{}c@{}}\$52.1k \\ (+$\,$9.8\%)\end{tabular} & 0.94 & \$51.0k & \begin{tabular}{@{}c@{}}$-$\$2.4k \\ (20.0 kW)\end{tabular} & \begin{tabular}{@{}c@{}}\$78.4k \\ (+$\,$11.2\%)\end{tabular} & 0.99  \\
  \hline
  \begin{tabular}{@{}c@{}}\textbf{W/o RH}\\ \textbf{0.5-hr PD price}\end{tabular} & 468 & \$32.9k & \begin{tabular}{@{}c@{}}$-$\$13.6k \\ (112.3 kW)\end{tabular} & \begin{tabular}{@{}c@{}}\$62.3k \\ (+$\,$31.1\%)\end{tabular} & 0.84 & \$45.6k & \begin{tabular}{@{}c@{}}$-$\$3.5k \\ (29.2 kW)\end{tabular} & \begin{tabular}{@{}c@{}}\$85.0k \\ (+$\,$20.6\%)\end{tabular} & 0.87  \\
  \hline
  \textbf{Coupled RH} & 486 & \$32.8k & \begin{tabular}{@{}c@{}}$-$\$14.0k \\ (115.5 kW)\end{tabular} & \begin{tabular}{@{}c@{}}\$63.4k \\ (+$\,$33.4\%)\end{tabular} & 0.83 & \$45.4k & \begin{tabular}{@{}c@{}}$-$\$3.8k \\ (31.2 kW)\end{tabular} & \begin{tabular}{@{}c@{}}\$86.2k \\ (+$\,$22.3\%)\end{tabular} & 0.85 \\
  \specialrule{.1em}{0em}{0em}
  \end{tabular}
  \vspace{-0.5em}
\end{table*}

The benefit of battery RHO is its adaptability to changes in forecasts over time. While the (forecast) PD prices, updated every 30 minutes and can be obtained from the Australian NEM \cite{aemowebsite}, variations in the consumption behaviour of end users can be obtained by solving the optimisation problem in \eqref{eqn:pro_sol}. Figures \ref{fig:look-ahead_price} and \ref{fig:look-ahead_cons} show changes in PD prices and aggregated expected consumption over time, respectively. To provide context, the values indicated by the blue line signify the forecast values obtained using 6 hours prior data; for example, the value at 20:00 was estimated at the receding horizon starting at 14:00. This dynamic behaviour in end-user consumption is achieved by incorporating changing PD prices, coupled with the application of behavioural economic concepts, including loss aversion and time inconsistency, as introduced in \cite{dinh2023modelling}. Furthermore, it can be seen that end users perform load shifting with relatively smooth curves, except for sudden jump at 20:00, right after DNSP peak hours window, which has a network usage charge of more than 27\textcent/kWh, as shown in Table \ref{tab:prosumer_charges}. This behaviour can be observed in real life, as reported by DNSP after one year of trial tariff \cite{Ausgridtrialtariff24}. 

\begin{table}[t]
    \vspace{-0.2 em}
  \caption{Pre-dispatch (PD) prices error analysis for in-sample period}
  \vspace{0 em}
  \label{tab:price_error}
  \centering
  \renewcommand{\arraystretch}{1.3}
  \begin{tabular}{c|ccccc}
  \specialrule{.15em}{0em}{0em}
  \textbf{$\sum_j (\lambda^\text{PD}_{j,h} - \lambda^\text{RT}_{j,h})$} & \textbf{Mean} & \textbf{Median} & \textbf{SD} & \textbf{Skew} & \textbf{Kurt} \\
  \specialrule{.08em}{0em}{0em}
  0.5-hr PD ($h=1$) & 0.082 & 0.00 & 1.167 & 9.96 & 144.54 \\
  32-hrs PD ($h=|H|$) & 0.121 & 0.00 & 1.350 & 8.85 & 103.04 \\
  All intervals & 0.132 & 0.00 & 1.414 & 8.68 & 94.39 \\
  \specialrule{.1em}{0em}{0em}
  \end{tabular}
  \vspace{-1.0 em}
\end{table}

\subsubsection{Optimal CBS capacities from different sizing methods}

Table \ref{tab:optimal_va} shows the optimal CBS capacity and the normalised annual total cost associated with each method. Note that after determining the optimal CBS capacity using various sizing methods, we run the CBS operation model in \eqref{eqn:CBS_operation} and calculate the cost using \eqref{eqn:ground_truth}. We also quantify the percentage of financial losses for the W/o RH and Coupled RH methods compared to the \textit{Exact} model, as shown by the percentage values in round brackets. Clearly, the Exact method provides the lowest cost in the in-sample period with a capacity of 250 kWh, which is obtained by exhaustively searching for the global optimal value. Nevertheless, the result for the Exact method is confirmed using the data from the out-of-sample, which also gives the lowest cost among the CBS capacities obtained.

Next, we see that the optimal solution obtained by the Coupled RH method returned the highest CBS capacity, resulting in the highest cost and financial losses. This can be attributed to the errors in the PD prices, as can be seen in the top panel of Fig. \ref{fig:look-ahead_price} and Table \ref{tab:price_error}. The analysis shows that the PD prices are more accurate if they are closer to RT, as shown by the lower mean and standard deviation (SD) in $h=1$ than those in $h=|H|$ and all intervals combined. Furthermore, high kurtosis indicates that, although most errors from PD prices cluster around the distribution mean, there are a few significant outliers. Mostly, these outliers lie on the right-hand side of the distribution mean, as indicated by the positive skew. Overall, this shows that PD prices generally overestimate true dispatch prices. As a result, both the W/o RH (which relies on the 0.5-hour PD prices) and the Coupled RH methods significantly oversize the CBS capacity, as both consider PD prices during the sizing process.

\begin{figure}[!t]
    \centerline{\includegraphics[trim={0.25cm 0cm 0.2cm 0.2cm}, clip, width=0.65\linewidth]{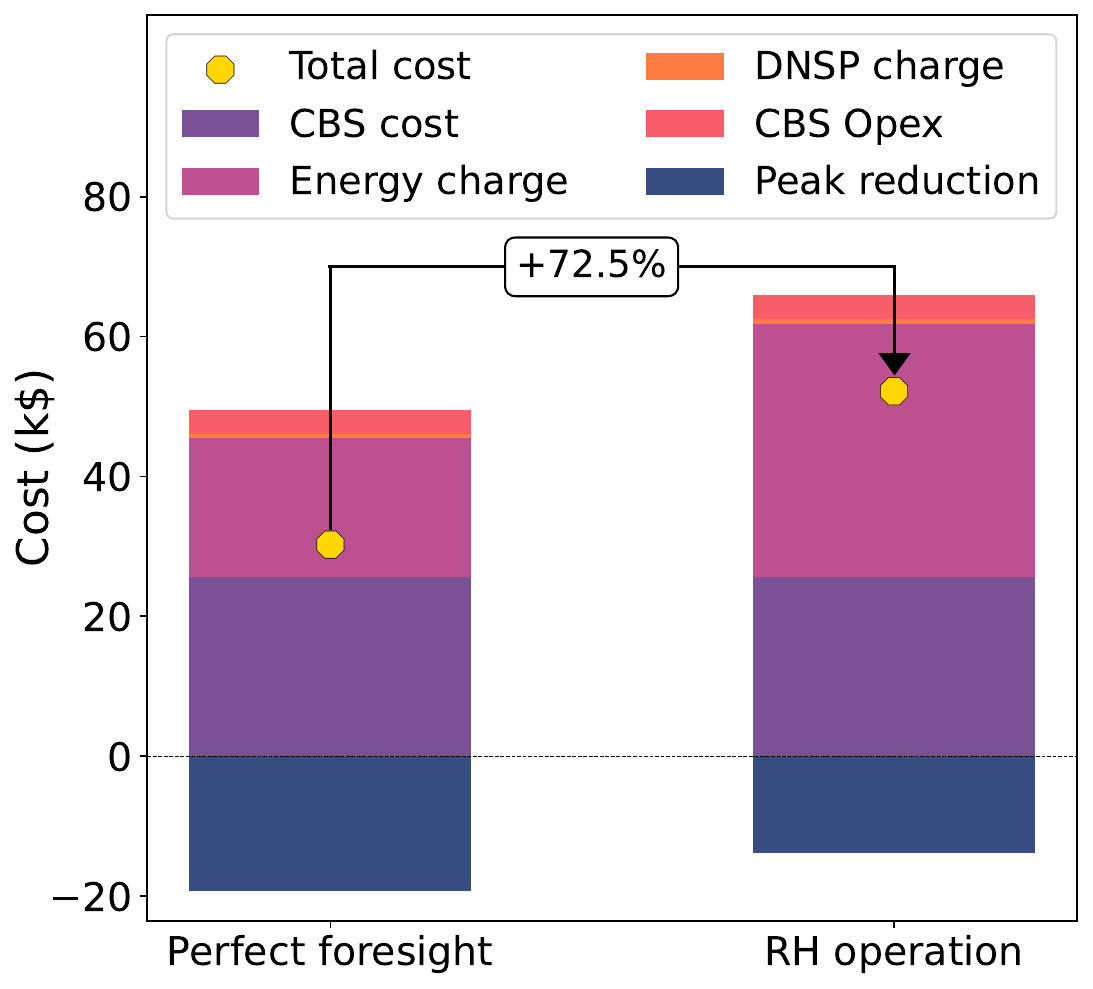}}
    \vspace{-0.2em}
    \caption{Perfect foresight vs RH operation with imperfect forecasts cost breakdown for 320 kWh CBS capacity in the in-sample period}
    \label{fig:perf_fore_comparison}
    \vspace{-1.0em}
\end{figure}

Although the W/o RH method with perfect foresight produces the storage size closest to the global optimal value, it still oversizes the CBS capacity, leading to approximately 9.8\% and 11.2\% financial losses in the in-sample and out-of-sample periods, respectively. This behaviour can be explained by the under-utilised CBS capacity in actual operation. Figure \ref{fig:perf_fore_comparison} compares the total costs of CBS capacity of 320 kWh between the perfect foresight and RHO with imperfect forecasts scenarios. 
Since the CBS capacity of 320 kWh is the optimal solution from the W/o RH method with perfect foresight, its cost breakdown can be obtained directly from the optimised solutions in \eqref{eqn:CBS_sizing_w/oRH}. As a result, it represents the minimum cost that a 320 kWh CBS can achieve. On the contrary, the cost breakdown for the RHO with imperfect forecasts is obtained by running the CBS operation model in \eqref{eqn:CBS_operation} for a 320 kWh CBS. It can be seen that the total cost difference lies mainly in the energy charge and revenue from peak demand reduction. While the energy charge depends heavily on the forecast of PD prices, the revenue from peak demand reduction depends greatly on the end-user (peak) consumption forecast. 

The under-utilisation of the CBS in the actual operation is also reflected in the daily average CBS cycle, as shown in Table \ref{tab:optimal_va}. Due to the imperfect forecast of price and consumption variations, CBS may overlook opportunities for energy arbitrage. This can lead to reduced daily charging and discharging activities, particularly when dealing with higher CBS capacities. In both the in-sample and out-of-sample periods, the battery size obtained from the Exact method experiences almost one cycle per day, which is the default warranty term across battery manufacturers. 

\subsubsection{Impact of CBS capacity on peak demand reduction}
Against the prevailing view, a higher CBS capacity does not guarantee a greater reduction in peak demand. It can be seen in the out-of-sample period in Table \ref{tab:optimal_va} that the lowest CBS capacity (250 kWh) gives the highest peak demand reduction (50.8 kW). As previously explained, the ability to reduce peak demand depends greatly on the end-user (peak) consumption forecast. It is shown in Fig. \ref{fig:look-ahead_cons} that predicted consumption (e.g., the blue line) could be higher than the actual (realised) consumption. Consequently, the CBS optimisation model miscalculates the actual peak demand and charges the CBS in earlier intervals, causing a new peak for the local neighbourhood. The higher the maximum CBS power, the higher the new peak, causing a lower reduction in peak demand. This issue has also been observed in real life in which a 1.1MW/2.15MWh battery in the distribution network caused a higher peak demand when performing energy arbitrage \cite{UQBattery}. Thus, it is crucial to ensure that CBS operates without detrimentally impacting distribution networks, and remains a vital area for research.

\section{Conclusion}

In this paper, we shed light on the impact of simplified models in state-of-the-art battery sizing studies, namely the W/o RH and Coupled RH approaches. To accurately quantify the financial losses from these simplifications, we developed a mathematical framework for a CBS-related business model using the trial tariffs from an Australian DNSP. We showed that the Coupled RH technique produced the least accurate results with significantly higher cost and CBS capacity than the Exact model, which considers the practical battery RHO. Although the W/o RH method, under perfect foresight, also resulted in an oversized battery capacity, the resulting financial losses were less significant. In addition, we highlighted a potential scenario in which CBS can negatively affect distribution networks by introducing new peak demand due to CBS arbitrage. In our future work, we want to focus on the use of CBS to provide other services for the electricity market and a profitable case study for CBS.

\bibliographystyle{IEEEtran}
\bibliography{reference.bib}

\begin{thebibliography}{10}
\providecommand{\url}[1]{#1}
\csname url@samestyle\endcsname
\providecommand{\newblock}{\relax}
\providecommand{\bibinfo}[2]{#2}
\providecommand{\BIBentrySTDinterwordspacing}{\spaceskip=0pt\relax}
\providecommand{\BIBentryALTinterwordstretchfactor}{4}
\providecommand{\BIBentryALTinterwordspacing}{\spaceskip=\fontdimen2\font plus
\BIBentryALTinterwordstretchfactor\fontdimen3\font minus
  \fontdimen4\font\relax}
\providecommand{\BIBforeignlanguage}[2]{{%
\expandafter\ifx\csname l@#1\endcsname\relax
\typeout{** WARNING: IEEEtran.bst: No hyphenation pattern has been}%
\typeout{** loaded for the language `#1'. Using the pattern for}%
\typeout{** the default language instead.}%
\else
\language=\csname l@#1\endcsname
\fi
#2}}
\providecommand{\BIBdecl}{\relax}
\BIBdecl

\bibitem{Alkimos}
\BIBentryALTinterwordspacing
Synergy. (2021) Alkimos beach energy storage trial final report. [Online].
  Available:
  \url{https://arena.gov.au/knowledge-bank/alkimos-beach-energy-storage-trial-final-report/}
\BIBentrySTDinterwordspacing

\bibitem{Yarra}
\BIBentryALTinterwordspacing
Yarra. (2022) Yarra community battery project. [Online]. Available:
  \url{https://www.yef.org.au/community-batteries/yarra-community-battery-trial/}
\BIBentrySTDinterwordspacing

\bibitem{AusgridCBS}
\BIBentryALTinterwordspacing
Ausgrid. (2022) Community batteries. [Online]. Available:
  \url{https://www.ausgrid.com.au/In-your-community/Community-Batteries}
\BIBentrySTDinterwordspacing

\bibitem{medi2020an}
C.~P. Mediwaththe, M.~Shaw, S.~Halgamuge, D.~B. Smith, and P.~Scott, ``An
  incentive-compatible energy trading framework for neighborhood area networks
  with shared energy storage,'' \emph{IEEE Transactions on Sustainable Energy},
  vol.~11, no.~1, pp. 467--476, 2020.

\bibitem{Ransan2021applying}
\BIBentryALTinterwordspacing
H.~Ransan-Cooper, B.~C.~P. Sturmberg, M.~E. Shaw, and L.~Blackhall, ``Applying
  responsible algorithm design to neighbourhood-scale batteries in australia,''
  \emph{Nature Energy}, vol.~6, no.~8, pp. 815--823, Aug 2021. [Online].
  Available: \url{https://doi.org/10.1038/s41560-021-00868-9}
\BIBentrySTDinterwordspacing

\bibitem{dinh2022optimal}
\BIBentryALTinterwordspacing
N.~T. Dinh, S.~A. Pourmousavi, S.~Karimi-Arpanahi, Y.~P.~S. Kumar, M.~Guo,
  D.~Abbott, and J.~A.~R. Liisberg, ``Optimal sizing and scheduling of
  community battery storage within a local market,'' in \emph{Proceedings of
  the Thirteenth ACM International Conference on Future Energy Systems}, ser.
  e-Energy '22.\hskip 1em plus 0.5em minus 0.4em\relax New York, NY, USA:
  Association for Computing Machinery, 2022, p. 34–46. [Online]. Available:
  \url{https://doi.org/10.1145/3538637.3538837}
\BIBentrySTDinterwordspacing

\bibitem{AERtrial}
\BIBentryALTinterwordspacing
A.~E. Regulator. (2022) Tariff trials. [Online]. Available:
  \url{https://www.aer.gov.au/networks-pipelines/network-tariff-reform/tariff-trials}
\BIBentrySTDinterwordspacing

\bibitem{Ausgridtrial}
\BIBentryALTinterwordspacing
Ausgrid. (2022) Trial tariffs network price list 2022-2023. [Online].
  Available:
  \url{https://cdn.ausgrid.com.au/-/media/Documents/Regulation/Pricing/PList/Ausgrids-Trial-Tariffs-Network-Price-List-2022-23.pdf}
\BIBentrySTDinterwordspacing

\bibitem{subthreshold}
\BIBentryALTinterwordspacing
------. (2022) Ausgrid sub-threshold tariffs 2022-23. [Online]. Available:
  \url{https://www.aer.gov.au/system/files/Ausgrid\%20-\%20Tariff\%20trial\%20notification\%20-\%202022-23\_0.pdf}
\BIBentrySTDinterwordspacing

\bibitem{khezri2022optimal}
\BIBentryALTinterwordspacing
R.~Khezri, A.~Mahmoudi, and H.~Aki, ``Optimal planning of solar photovoltaic
  and battery storage systems for grid-connected residential sector: Review,
  challenges and new perspectives,'' \emph{Renewable and Sustainable Energy
  Reviews}, vol. 153, p. 111763, 2022. [Online]. Available:
  \url{https://www.sciencedirect.com/science/article/pii/S1364032121010339}
\BIBentrySTDinterwordspacing

\bibitem{kani2020improving}
S.~A.~P. Kani, P.~Wild, and T.~K. Saha, ``Improving predictability of renewable
  generation through optimal battery sizing,'' \emph{IEEE Transactions on
  Sustainable Energy}, vol.~11, no.~1, pp. 37--47, 2020.

\bibitem{tilman2021renewable}
\BIBentryALTinterwordspacing
T.~Weckesser, D.~F. Dominković, E.~M. Blomgren, A.~Schledorn, and H.~Madsen,
  ``Renewable energy communities: Optimal sizing and distribution grid impact
  of photo-voltaics and battery storage,'' \emph{Applied Energy}, vol. 301, p.
  117408, 2021. [Online]. Available:
  \url{https://www.sciencedirect.com/science/article/pii/S0306261921008059}
\BIBentrySTDinterwordspacing

\bibitem{da2022a}
\BIBentryALTinterwordspacing
D.~Huo, M.~Santos, I.~Sarantakos, M.~Resch, N.~Wade, and D.~Greenwood, ``A
  reliability-aware chance-constrained battery sizing method for island
  microgrid,'' \emph{Energy}, vol. 251, p. 123978, 2022. [Online]. Available:
  \url{https://www.sciencedirect.com/science/article/pii/S0360544222008817}
\BIBentrySTDinterwordspacing

\bibitem{khezri2021a}
R.~Khezri, A.~Mahmoudi, and M.~H. Haque, ``A demand side management approach
  for optimal sizing of standalone renewable-battery systems,'' \emph{IEEE
  Transactions on Sustainable Energy}, vol.~12, no.~4, pp. 2184--2194, 2021.

\bibitem{baker2017energy}
K.~Baker, G.~Hug, and X.~Li, ``Energy storage sizing taking into account
  forecast uncertainties and receding horizon operation,'' \emph{IEEE
  Transactions on Sustainable Energy}, vol.~8, no.~1, pp. 331--340, 2017.

\bibitem{taheri2019energy}
S.~Taheri, V.~Kekatos, and S.~Veeramachaneni, ``Energy storage sizing in
  presence of uncertainty,'' in \emph{2019 IEEE Power \& Energy Society General
  Meeting (PESGM)}, 2019, pp. 1--5.

\bibitem{moghaddam2019optimal}
I.~N. Moghaddam, B.~Chowdhury, and M.~Doostan, ``Optimal sizing and operation
  of battery energy storage systems connected to wind farms participating in
  electricity markets,'' \emph{IEEE Transactions on Sustainable Energy},
  vol.~10, no.~3, pp. 1184--1193, 2019.

\bibitem{AmberElectric}
\BIBentryALTinterwordspacing
Amber. (2022) Amber electric. [Online]. Available:
  \url{https://www.amber.com.au/}
\BIBentrySTDinterwordspacing

\bibitem{smart2022the}
\BIBentryALTinterwordspacing
T.~Conversation. (2022) Smart meters and dynamic pricing can help consumers use
  electricity when it’s less costly, saving money and reducing pollution.
  [Online]. Available:
  \url{https://theconversation.com/smart-meters-and-dynamic-pricing-can-help-consumers-use-electricity-when-its-less-costly-saving-money-and-reducing-pollution-190217}
\BIBentrySTDinterwordspacing

\bibitem{dinh2023modelling}
N.~T. Dinh, S.~Karimi-Arpanahi, S.~A. Pourmousavi, R.~Yuan, M.~Guo, J.~A.~R.
  Liisberg, and J.~Lemos-Vinascco, ``Modelling irrational behaviour of
  residential end users using non-stationary gaussian processes,'' 2023.

\bibitem{aemowebsite}
\BIBentryALTinterwordspacing
AEMO. (2022) Aemo market portals. [Online]. Available:
  \url{https://aemo.com.au/}
\BIBentrySTDinterwordspacing

\bibitem{werner2021pricing}
\BIBentryALTinterwordspacing
L.~Werner, A.~Wierman, and S.~H. Low, ``Pricing flexibility of shiftable demand
  in electricity markets,'' in \emph{Proceedings of the Twelfth ACM
  International Conference on Future Energy Systems}, ser. e-Energy '21.\hskip
  1em plus 0.5em minus 0.4em\relax New York, NY, USA: Association for Computing
  Machinery, 2021, p. 1–14. [Online]. Available:
  \url{https://doi.org/10.1145/3447555.3464847}
\BIBentrySTDinterwordspacing

\bibitem{aemofactsheet}
\BIBentryALTinterwordspacing
AEMO. (2023) The national electricity market. [Online]. Available:
  \url{https://aemo.com.au/-/media/files/electricity/nem/national-electricity-market-fact-sheet.pdf}
\BIBentrySTDinterwordspacing

\bibitem{Ausgrid2012data}
\BIBentryALTinterwordspacing
Ausgrid. (2012) Solar home electricity data. [Online]. Available:
  \url{https://www.ausgrid.com.au/Industry/Our-Research/Data-to-share/Solar-home-electricity-data}
\BIBentrySTDinterwordspacing

\bibitem{AEC2021}
\BIBentryALTinterwordspacing
A.~E. Council, ``Solar report quarter 3, 2021,'' Tech. Rep., 2021. [Online].
  Available:
  \url{https://www.energycouncil.com.au/media/5zylveyr/australian-energy-council-solar-report\_q3-2021.pdf}
\BIBentrySTDinterwordspacing

\bibitem{githubproject}
\BIBentryALTinterwordspacing
N.~T. Dinh. (2023) Github project. [Online]. Available:
  \url{https://github.com/nam-dinh-codes/community-battery-sizing-study}
\BIBentrySTDinterwordspacing

\bibitem{Ausgridtrialtariff24}
\BIBentryALTinterwordspacing
Ausgrid. (2023) 2023-24 sub-threshold tariff notification. [Online]. Available:
  \url{https://www.aer.gov.au/system/files/Ausgrid\%20-\%20Tariff\%20trial\%20notification\%20-\%202023-24.pdf}
\BIBentrySTDinterwordspacing

\bibitem{UQBattery}
\BIBentryALTinterwordspacing
T.~U. of~Queensland. (2020) The business case for behind-the-meter energy
  storage. [Online]. Available:
  \url{https://sustainability.uq.edu.au/files/11868/EPBQtyRptq12020.pdf}
\BIBentrySTDinterwordspacing

\end{thebibliography}



%



\end{document}